\documentstyle[12pt]{article}
\newlength{\extraspace}
\newlength{\extraspaces}
\newcommand{\bq}{\begin{eqnarray}
 \addtolength{\abovedisplayskip}{\extraspaces}
 \addtolength{\belowdisplayskip}{\extraspaces}
 \addtolength{\abovedisplayshortskip}{\extraspace}
 \addtolength{\belowdisplayshortskip}{\extraspace}}
\newcommand{\eq}{\end{eqnarray}}

\newcommand{\newsection}[1]
 {\vspace{5mm}
 \pagebreak[3]
 \addtocounter{section}{1}
 \setcounter{equation}{0}
 \setcounter{subsection}{0}
 \setcounter{footnote}{0}
 \begin{flushleft}
 {\large\bf \thesection. #1}
 \end{flushleft}
 \nopagebreak
 \medskip
 \nopagebreak}

\begin{document}

\addtolength{\baselineskip}{.8mm}

\thispagestyle{empty}

\begin{flushright}
{\sc OUTP-95-26P}\\
 Oct 1995\\
 hep-th/9506190
\end{flushright}
\vspace{.3cm}

\begin{center}
{\large\sc{Topologically Massive Gauge Theory with
 $O(2)$ Symmetry} }\\[15mm]

{\sc  Ian I. Kogan}\footnote{On leave of absence from
  ITEP, Moscow, 117259, Russia} and {\sc Kai-Ming Lee}\\[2mm]
{\it Theoretical Physics, University of Oxford,\\
 1 Keble Road, Oxford, OX1 3NP, UK}\\[15mm]
{\sc Abstract}
\end{center}

\noindent
We discuss the structure of the vacua in $O(2)$ topologically
massive gauge theory on a torus. Since $O(2)$ has two connected
components, there are four classical vacua. The different
vacua impose different boundary conditions on the gauge potentials.
We also discuss the non-perturbative transitions between the
vacua induced by vortices of the theory.\\
11.15.Ex, 11.15.Kc.

\vfill
\newpage
\pagestyle{plain}
\setcounter{page}{1}
\stepcounter{subsection}
\newsection{Introduction}
\renewcommand{\footnotesize}{\small}

Topologically massive gauge theories (TMGT) \cite{tmgt} are of great
interest for many years. They have important applications
in different areas of theoretical  and mathematical physics,
for example quantum Hall effect \cite{qhe}, knot invariants and
conformal field theories \cite{witten} and many other areas.

The structure of the Hilbert space in these theories may be rather
unusual. For example, in abelian $U(1)$ TMGT one can see \cite{km}
that the  Hilbert space of the theory is a direct  product of the
massive gauge particle Hilbert space and  some quantum-mechanical
Hilbert space which  is the product of $g$ copies (for a genus $g$
Riemann surface) of the Hilbert space for the  Landau problem on the
torus. In the infinite mass limit all levels except the first one
are  decoupled  as well as the massive particle Hilbert space and
we then have only the first Landau level which becomes the Hilbert
space of the topological  Chern-Simons theory.

In this letter we shall discuss some unusual properties of
the $O(2)$ theory, which can be obtained after a spontaneous
breaking of the $SU(2)$ or $SO(3)$ symmetry by a Higgs field in
a $5$-dimensional representation \cite{otwo}. We shall see that
because $O(2)$, contrary to $U(1)$, has two connected components,
i.e. $\pi_0(O(2)) = Z_2$, one shall get new classical vacua in
$O(2)$ theory. We shall not only consider all these vacua but
also find the particle spectra corresponding to each vacuum.

Let us note that the problem of this type in pure Chern-Simons
theory was discussed bye Moore and Seiberg \cite{ms}. They
could study only the properties of the ground state, because
there are no excitations in the pure Chern-Simons theory. Our
results for the ground state (obtained in a full TMGT and using
other methods than in a pure Chern-Simons theory) are in
agreement with the results for pure Chern-Simons theory, as
it must be.

We shall start with a brief review of the ordinary $U(1)$ TMGT.
Then we shall go to the $O(2)$ model, which will be defined as
a low-energy limit of the $SO(3)$ theory. We will also discuss
the non-perturbative transitions between different vacua in the
full $SO(3)$ theory. These transitions are induced by the
instantons, which in this case, are the vortices associated with
the non-trivial component of the $O(2)$.

\newsection{Abelian Topologically Massive Gauge Theory}

Let us consider an  abelian topologically massive gauge theory \cite{tmgt}:
\begin{equation}
{\cal L}_{U(1)} =
 -\frac{1}{4\gamma}F_{\mu\nu}F^{\mu\nu} +
\frac{k}{8\pi} \epsilon_{\mu\nu\lambda}
A_{\mu}\partial_{\nu}A_{\lambda} .
\label{eq:tmgt}
\end{equation}
We assume that the spacetime is in the form $M\times R$ where
$M$ is a Riemann surface. We choose the $A_0=0$ gauge to perform
the canonical quantization. The corresponding constraint is
\begin{equation}
\frac{1}{\gamma}\partial_i\dot A_i +
\frac{k}{8\pi} \epsilon_{ij}F_{ij} =0. \label{eq:constraint}
\end{equation}
The vector potential on a plane can be represented as
$A_i=\partial_i\xi+\epsilon_{ij}\partial_j\chi$,
substitute it into the constraint, one gets
$\partial^2\dot\xi=(k\gamma/4\pi)\partial^2\chi$.
Forget the zero modes for the moment, this is
$\dot\xi=(k\gamma/4\pi)\chi=M\chi$.
Substituting this constraint into the Lagrangian (\ref{eq:tmgt})
one gets
\begin{equation}
{\cal L}=\frac{1}{2\gamma} ((\partial_i\dot\chi)^2-
(\partial^2\chi)^2 - M^2\chi\partial^2\chi)
\end{equation}
which is a free Lagrangian for the field
$\Phi=\sqrt{\partial^2/ \gamma} \chi$
\begin{equation}
{\cal L}=\frac{1}{2} (\dot{\Phi}^2-(\partial_i\Phi)^2-M^2\Phi^2)
\end{equation}
describing the free particle with mass $M =\gamma k/4\pi$.

It is easy to see that on the plane the spatial independent
fields $A_i(x,t) = {\bf A}_i(t)$ also satisfy the constraint.
For these fields one gets the Landau Lagrangian \cite{km}
\begin{equation}
{\cal L} = \frac{1}{2\gamma}\dot{\bf A}_i^2-\frac{k}{8\pi}
  \epsilon_{ij}{\bf A}_i\dot{\bf A}_j\label{eq:Landau}
\end{equation}
which describes the particle with mass $m = \gamma^{-1}$ on
the plane ${\bf A}_x,{\bf A}_y$ in a magnetic field $B = k/4\pi$.
The mass gap is $M = B/m = \gamma k/4\pi$ which is precisely the
mass of the gauge particle.

Let us note that  ${\bf A}_x$ and ${\bf A}_y$ belong to the
configuration space, however if reduced to  the first Landau
level, which means $ m = 1/\gamma \to 0$, the theory reduces
to the  pure Chern-Simons theory which is an exactly solvable
$2+1$ dimensional topological field theory.

For general 2-dimensional Riemann surface of genus $g$, any
one-form $A$ can be uniquely decomposed  according to Hodge
theorem as
\begin{eqnarray}
 A=d\xi+\delta\chi+{\bf A},~~~~~~~~~~~d{\bf A}=\delta{\bf A}=0
\end{eqnarray}
which generalizes the decomposition on the plane we have used
before. The harmonic form ${\bf A}$ equals
\begin{eqnarray}
{\bf A}=\sum_{p=1}^g({\bf A}^p\alpha_p+{\bf B}^p\beta_p)
\end{eqnarray}
where $\alpha_p$ and $\beta_p$ are canonical harmonic 1-forms
($1$-cohomology) on a Riemann surface and there are precisely
$2g$ harmonic 1-forms  on genus $g$ Riemann surface (two in case
of a torus). After diagonalization one finds that there are $g$
copies of the Landau problem and the total Hilbert space $\cal H$
of the abelian topologically massive gauge theory
\begin{equation}
{\cal H} = {\cal H}_\Phi \otimes\prod_{i=1}^g
{\cal H}_{\bf A}
\end{equation}
is the product of the free massive particle Hilbert space
${\cal H}_\Phi$ and $g$ copies of the Landau problem's
Hilbert space ${\cal H}_{\bf A}$.

Let us concentrate on the case of a torus. We get the Landau
problem on the plane $({\bf A}_x,{\bf A}_y)$. We must not
forget about large gauge transformations acting on the
quantum-mechanical coordinates ${\bf A}_i\rightarrow{\bf A}_i
+2\pi N_i$, where $N_i$ are integers. These transformations act
on gauge potential because the only gauge-invariant objects one
can construct for ${\bf A}_i$ -- Wilson lines
\begin{equation}
 W(C) = exp(i\oint_C A_\mu dx^\mu)
\end{equation}
are invariant under these transformations (we choose coordinate
on a torus in a way that $x\sim x+1$ and $y\sim y+1$) and one
can consider torus $0\leq{\bf A}_i<2\pi$ with the area $(2\pi)^2$.

Let us note that being reduced to the first Landau level this
torus becomes the phase space - thus for the consistent
quantization this area must be proportional to the integer
(the total number of the states must be integer). It is known
that the density of states $\rho$  on Landau level equals to
$B/2\pi$, where $B$ is a magnetic field. In our case the ``magnetic
field'' in $({\bf A}_x,{\bf A}_y)$ plane is equals to $B = (k/4\pi)$,
thus the total number of states will be $N=(1/2\pi)
(k/4\pi)\times(2\pi)^2=k/2$. Thus, $k$ must be an even integer.
In more general case we have to enlarge our phase space
to have minimal possible integer number of states (for rational $k$)
or even infinite number (for irrational $k$).

\newsection{The $O(2)$ Model}
 From now on, we turn to the TMGT with $O(2)$ symmetry on a torus.
We obtain the gauge group $O(2)$ by spontaneous breaking
a non-abelian Higgs model with Chern-Simons term and gauge group
$SU(2)$ or $SO(3)$. We usually suppose that the symmetry breaking
scale is very large and just consider the low energy phenomena.
The reduced system is a TMGT with $O(2)$ symmetry.

The full Lagrangian is
\begin{equation}
{\cal L}=\frac{1}{8\gamma} {\rm Tr} F_{\mu\nu}F^{\mu\nu} +
  \frac{k}{16\pi} {\rm Tr} (A\,dA +\frac{2}{3}A^3)
  +\frac{1}{4} {\rm Tr}(\partial \phi +[A,\phi])^2
  -V(\phi).
\label{Lagrangian}\end{equation}
The normalization is chosen such that it will agree with the
abelian theory in some special case. We choose the representation
and the potential of $\phi$ such that the unbroken group is $O(2)$.
The simplest choice is the following \cite{otwo}. The Higgs field is a
$3\times 3$ real symmetric traceless matrix. If $g \in SO(3)$,
the action of $g$ on $\phi$ is
\begin{equation}
g(\phi)= g \phi g^{-1}.
\end{equation}
The potential of the Higgs field is
\begin{equation}
V(\phi)=\lambda_1({\rm Tr}\phi^2 -6v^2)^2+\lambda_2(\det\phi+2v^3)^2
\label{potentialofHiggs}\end{equation}
where $\lambda_1$ is of dimension of mass and $\lambda_2$ is
dimensionless. They are both positive. $V(\phi)$ is invariant
under $SO(3)$ and its value is zero if and only if the eigenvalues
of $\phi$ are $1$, $1$ and $-2$. If the vacuum expectation value
of $\phi$ is
\begin{equation}
\langle \phi\rangle=v
\left( \begin{array}{ccc} 1&0&0\\ 0&1&0\\ 0&0&-2\end{array}\right),
\label{vev}\end{equation}
the group $O(2)$ embedded in the $SO(3)$ is then
\begin{equation}
O(2) =\{\left( \begin{array}{ccc} \cos\theta&\sin\theta&0\\
                                 -\sin\theta&\cos\theta&0\\
                                  0&0&1\end{array}\right),
        \left( \begin{array}{ccc} \cos\theta&-\sin\theta&0\\
                                 -\sin\theta&-\cos\theta&0\\
                                  0&0&-1\end{array}\right)\}.
\end{equation}
Abstractly, $O(2)$ could be described as an $U(1)$ with an extra
element $X$, where $e^{i\theta} X = X e^{-i \theta}$. In the above
representation, $X$ is $\left(\begin{array}{ccc}1&0&0\\0&-1&0\\
0&0&-1\end{array}\right)$.

We will usually assume that the Higgs field takes its vacuum
expectation values and drop the terms independent of the gauge
field from the Lagrangian.

\newsection{Topological Considerations}
If we ignore the Higgs field and take the limit $\gamma\to\infty$,
$\cal L$ goes to the pure Chern-Simons Lagrangian. The classical
solutions of the pure Chern-Simons theory are given by
\begin{equation}
{\rm Hom}(\pi_1(M),G)/G
\end{equation}
where $G$ is the gauge group and $M$ is the space manifold.

If $M$ is a torus, $\pi_1(M)$ is $Z\times Z$, the direct product
of two copies of the set of all integers. We can label
the classical solutions by a pair $(g_1,g_2)$ where $g_1$ and
$g_2$ commute in $G$ and we identify $(g_1,g_2)$ with
$(g g_1 g^{-1}, g g_2 g^{-1})$. The pair describes the
holonomy of the two non-trivial loops on the torus.

For $O(2)$, the solutions are one continuous family
$(e^{i\theta_1},e^{i\theta_2})$, which identifies with
$(e^{-i\theta_1},e^{-i\theta_2})$, and three discrete ones
$(X,\pm 1)$, $(\pm 1, X)$ and $(X, \pm X)$. We will see that
the sector corresponding to the continuous family has zero modes
but the sector corresponding to the discrete solutions have none.

If we consider that the $O(2)$ group is embedded in $SO(3)$,
the fundamental group of the vacuum manifold is
$\pi_1(SO(3)/O(2)) = \pi_0(O(2))=Z_2$. Therefore, there are
point-like topological excitations, vortices, in our theory.
In $3+1$ dimensional spacetime, they are the cosmic strings. We will
see that a pair of this vortex-anti-vortex is the instanton that
mixes different classical vacua.

The second homotopy group of the vacuum manifold also has
physical effect. The generator of $\pi_2(SO(3)/O(2)) = \pi_1(O(2))=Z$
is the magnetic monopole in $3+1$ dimensional spacetime. In this $O(2)$
theory, the monopole induces the large gauge transformation
discussed in Section 2.

\newsection{The Four Sectors}
We are going to solve the equations of motion derived from
(\ref{Lagrangian}) and find out the classical vacua.
Define a basis of $SO(3)$ by
$T^1=\left(\begin{array}{ccc}0&1&0\\-1&0&0\\0&0&0\end{array}\right)$,
$T^2=\left(\begin{array}{ccc}0&0&-1\\0&0&0\\1&0&0\end{array}\right)$ and
$T^3=\left(\begin{array}{ccc}0&0&0\\0&0&1\\0&-1&0\end{array}\right)$.
Then, $[T^a,T^b]=\epsilon^{abc}T^c$. The gauge potential is
\begin{equation}
A=\left(\begin{array}{ccc}
0&A^1&-A^2\\-A^1&0&A^3\\A^2&-A^3&0\end{array}\right).
\end{equation}
We choose the $A_0=0$ gauge. The equation of motion of the spatial
components of the gauge potential is
\begin{equation}
\frac{1}{\gamma}(\partial_j F_{ji} +[A_j,F_{ji}])^a
 - \frac{1}{2}{\rm Tr}((\partial_i\phi +[A_i,\phi])[T^a,\phi])=0,
\end{equation}
the equation of motion of the Higgs field is in the form
\begin{equation}
D_\mu D^\mu \phi - V'(\phi)=0
\label{eofmHiggs}\end{equation}
where $D_\mu$ is the covariant derivative. In addition to these
two equations of motion, we also have the constraint
\begin{equation}
\frac{1}{\gamma}(\partial_i\dot A^a_i +\epsilon^{abc}A^b_i
\dot A^c_i) +\frac{k}{8\pi}\epsilon^{ij}F^a_{ij}
+\frac{1}{2}{\rm Tr}(\partial_t\phi [T^a,\phi])=0.
\label{fullconstraint}\end{equation}

We consider only the low energy phenomena in this section.
Equivalently, we consider the limit where $\lambda_1$, $\lambda_2$
and $v$ of (\ref{potentialofHiggs}) go to infinity. In this limit,
the two equations of motion split into four. For example, when
$\lambda_1$ and $\lambda_2$ go to infinity, the derivative term
in (\ref{eofmHiggs}) is negligible and we have
\begin{equation}
V'(\phi)=0.
\label{vprime}\end{equation}
This in turn implies that
\begin{equation}
D_\mu D^\mu \phi=0.
\end{equation}
Similarly, when $v$ goes to infinity, we have
\begin{equation}
{\rm Tr}((\partial_i\phi +[A_i,\phi])[T^a,\phi])=0
\end{equation}
and
\begin{equation}
\frac{1}{\gamma}(\partial_j F_{ji} +[A_j,F_{ji}])^a=0.
\label{dfequzero}\end{equation}

One solution is that the value of the Higgs field is constant
in spacetime and the gauge potential is zero up to gauge transformation.
This is the trivial classical vacuum. We say that all the
excitations based on this vacuum are in the trivial
sector. By a suitable gauge transformation, we can assume that
the Higgs field is given by (\ref{vev}).

To find out the spectrum of the trivial sector, notice that the
interaction term of the Higgs and gauge potential in the
Lagrangian is, in this case,
\begin{equation}
{\rm Tr}(\partial\phi +[A,\phi])^2={\rm Tr}[A,\phi]^2
=18 v^2((A^2)^2+(A^3)^2).
\end{equation}
To consider only low energy phenomena, we set $A^2=A^3=0$.
The theory is reduced to the abelian TMGT. However, we also have
to identify states related by gauge transformations. Now, we have
one more element $X$ than the $U(1)$ case. Its action is to
identify $A^1$ with $-A^1$. Thus, we have only half of the states
as in the abelian case. The classical configuration space could
be chosen as $0\leq A^1_x <1$ and $0\leq A^1_y <\frac{1}{2}$. The
total number of quantum states in the first Landau level will be
$k/4$. Thus, to have a sensible quantum theory, $k$ must be a
multiple of $4$. In other cases we again must enlarge our phase
space before quantization in the same way as in the $U(1)$ case.
The Hilbert space structure is similar to that of the abelian theory.

We now consider the case that the holonomy of at least one of the
non-trivial loops is in the component of $O(2)$ other than the
identity component. We call it the non-trivial sector. We already
know that there are three non-trivial sectors for the $O(2)$ model
on a torus.

Let $X(\theta)=\left(\begin{array}{ccc}
1&0&0\\0&\cos\theta&\sin\theta\\0&-\sin\theta&\cos\theta
\end{array}\right)$ where $\theta$ would be chosen such that,
say, if we want to have non-trivial holonomy along the
$x$-direction, $\theta(x+L,y)=\theta(x,y)+\pi$ and
$\theta(x,y+L)=\theta(x,y)$. Other than this condition,
$\theta$ could be any function independent of time. We will
see that the low energy physics depends only on the boundary condition
stated above and not on the detail form of $\theta$. Note that
$X(\pi)=X$.

We will reinstall explicitly the dependence on the size of
torus. We choose the torus to be the same size, $L$, in both
directions. It is easy to generalize to tori of other sizes.

Now, we consider the following Higgs field and gauge potential:
\begin{eqnarray}
\phi &=v& X(\theta)
\left( \begin{array}{ccc} 1&0&0\\ 0&1&0\\ 0&0&-2\end{array}\right)
X(\theta)^{-1}\nonumber\\
&=v&
\left(\begin{array}{ccc}1&0&0\\
0&\cos^2\theta-2\sin^2\theta&-3\sin\theta\cos\theta\\
0&-3\sin\theta\cos\theta&\sin^2\theta-2\cos^2\theta\end{array}\right)
\label{vacHiggs}\end{eqnarray}
and
\begin{equation}
A=- \partial X X^{-1}=- \partial\theta T^3.
\label{vacgauge}\end{equation}
If $X$ was a single valued function on the torus, then these
Higgs field and gauge potential would be just the gauge transformation
of the corresponding vacuum fields in the trivial sector.
Although $X$ is not single valued, these Higgs field and
gauge potential are single valued and well defined on the torus.
It is now easy to see that they satisfy equations (\ref{vprime}) -
(\ref{dfequzero}) and the constraint (\ref{fullconstraint})
because those equations are gauge invariant.

Thus, the Higgs field and gauge potential in (\ref{vacHiggs})
and (\ref{vacgauge}) define the classical vacua in the
non-trivial sectors. (We can, of course, choose $\theta$ to
be zero and we obtain the vacuum of the trivial sector.)
We can choose any other values of the Higgs fields in the
vacuum manifold as long as they give the same holonomy. Any
such value of Higgs field will be gauge equivalent to this one.

We now calculate the potential of the gauge potential induced by
the Higgs field. Since
\begin{equation}
X(\theta)^{-1}AX(\theta)=
\left(\begin{array}{ccc}
  0&\cos\theta A^1+\sin\theta A^2&\sin\theta A^1-\cos\theta A^2\\
  -\cos\theta A^1-\sin\theta A^2&0&A^3\\
  \cos\theta A^2-\sin\theta A^1 &-A^3&0\end{array}\right),
\end{equation}
\begin{equation}\begin{array}{cc}
{\rm Tr}[A,\phi]^2&=v^2{\rm Tr}[X(\theta)^{-1}AX(\theta),
\left( \begin{array}{ccc} 1&0&0\\ 0&1&0\\ 0&0&-2\end{array}\right)
]^2\\
&=18 v^2 ((A^1 \sin\theta-A^2\cos\theta)^2 +(A^3)^2),
\end{array}\end{equation}
and
\begin{equation}
2 {\rm Tr}\partial \phi[A,\phi]=36 v^2 A^3\partial\theta.
\end{equation}
The effective potential of $A$ is
\begin{equation}
{\rm Tr}(\partial\phi+[A,\phi])^2=
18 v^2 ((A^1 \sin\theta-A^2\cos\theta)^2 + (A^3 +\partial\theta)^2).
\end{equation}
If we write the general gauge potential as the sum of the vacuum
value and the fluctuations, we see that in the large $v$ limit,
$A^3$ must be frozen to the vacuum value $-\partial\theta$ and
the fluctuations of the first two components of the gauge potential
must satisfy $A^1 \sin\theta=A^2\cos\theta$. Let $A^1=A \cos\theta$
and $A^2=A\sin\theta$. Here, $A$ is a vector field with
three components $(A_t,A_x,A_y)$ and not the Lie-algebra valued
one form in the previous equations. Note $A^1$ and $A^2$ are periodic
in $x$ and $y$ but $A$ is not periodic in the non-trivial sectors.
Thus $A$ does not have the corresponding zero modes.

The components of the gauge field can be expressed by $A$.
Define $F_{ij}=\partial_iA_j-\partial_jA_i$, we have
\begin{eqnarray}
F^1_{ij}&=&\partial_iA^1_j-\partial_jA^1_i
          +A^2_iA^3_j-A^2_jA^3_i\nonumber\\
&=&F_{ij}\cos\theta,\\
F^2_{ij}&=&F_{ij}\sin\theta,\\
F^3_{ij}&=&0.
\end{eqnarray}

The Lagrangian, which was
\begin{eqnarray}
{\cal L}&=&\frac{1}{2\gamma}(\dot A^a_i -\partial_i A^a_t
  +\epsilon^{abc}A^b_tA^c_i)^2
  -\frac{1}{4\gamma}F^a_{ij}F^a_{ij}
  -\frac{k}{8\pi}(\epsilon^{ij}A^a_i\dot A^a_j -A^a_t\epsilon^{ij}
  F^a_{ij}) \nonumber \\
  & &+ \frac{1}{4}{\rm Tr}(\partial\phi+[A,\phi])^2 - V(\phi),
\end{eqnarray}
is now, after neglecting the terms independent of $A$ and in the
$A_t^a=0$ gauge,
\begin{equation}
{\cal L}=\frac{1}{2\gamma}\dot A_i^2-\frac{1}{4\gamma}F_{ij}F_{ij}
  -\frac{k}{8\pi}\epsilon^{ij}A_i\dot A_j.\label{reducedLag}
\end{equation}
The constraint of the full system is
\begin{equation}
\frac{1}{\gamma}(\partial_i\dot A^a_i +\epsilon^{abc}A^b_i
\dot A^c_i) +\frac{k}{8\pi}\epsilon^{ij}F^a_{ij}=0.
\end{equation}
Express $A^a_i$ in terms of $A_i$ and $\theta$, this is
equivalent to
\begin{equation}
\frac{1}{\gamma}\partial_i\dot A_i
  +\frac{k}{8\pi}\epsilon^{ij}F_{ij}=0.\label{constraintone}
\end{equation}

Notice that the form of the Lagrangian and the constraint are
exactly the same as the abelian theory or the trivial sector.
However, recall that the boundary condition is that $A_i\cos\theta$
and $A_i\sin\theta$ are periodic. If the holonomy of the
$x$-direction is non-trivial, the effective gauge potential
must be anti-periodic in $x$. This distinguishes the non-trivial
sectors from the trivial sector. We also see that $\theta$ does not
appear in the Lagrangian or the constraint. It only imposes
the boundary condition on the effective gauge potential. Hence,
the detail form of $\theta$ is irrelevant.

Following the analysis of the $U(1)$ theory, it is easy to see
that the non-trivial sectors can also be described by a free
particle of mass $k\gamma /4\pi$. Because of the finite size of
the torus, the momentum of the particle is quantized. To satisfy
the anti-periodic boundary condition, only the ``half integer
modes'' are allowed in the direction where the holonomy is
non-trivial.

\newsection{Non-perturbative Transitions Between Sectors}
If we consider the full $SU(2)$ or $SO(3)$ theory, there will
be transitions between different sectors. Indeed, now one can
easily connect two different topological sectors by some field
configuration - but to do this it is necessary to excite some
heavy degrees of freedom. The instanton is one of these heavy
degrees of freedom which induces the transitions. The exact
instanton equation is non-linear and difficult to solve but we
could easily give a qualitative picture of what will happen.

Since $O(2)$ has two connected components, there will be
point-like topological defects, the vortices, in $2+1$
dimensional spacetime \cite{otwo}. It is, in fact, the same
reason why there are non-trivial sectors in the theory. We
will see that the instanton consists of a pair of
vortex-anti-vortex winding around a non-trivial loop on
the torus.

An ansatz of the vortex solution is the following: in the
polar coordinates $(r,\theta)$ on a plane,
\begin{equation}
\phi(r,\theta)=v(r)
\left(\begin{array}{ccc}1&0&0\\
0&\cos^2\frac{\theta}{2}-2\sin^2\frac{\theta}{2}&-3
  \sin\frac{\theta}{2}\cos\frac{\theta}{2}\\
0&-3\sin\frac{\theta}{2}\cos\frac{\theta}{2}&\sin^2\frac{\theta}{2}
  -2\cos^2\frac{\theta}{2}\end{array}\right)
\end{equation}
where $v(r)$ is a function of $r$ such that $v(0)=0$ and
$v(\infty)=v$. The gauge potential is such that
$A(0,\theta)=0$ and $A(\infty,\theta)=- T^3\,\frac{{\rm d}\theta}{2}$.
Comparing to (\ref{vacHiggs}) and (\ref{vacgauge}), we see
that at infinity, the Higgs field and the gauge potential
 approach their  vacuum  values.
  At  the  the origin  there are heavy excitations  in a region
  with size of order $1/v^2$ and  the mass of the vortex is of
order of magnitude $v^2$. The functional form of $v(r)$ and the
gauge potential could be determined by substituting them
into (\ref{Lagrangian}) and solving the differential
equations derived from it. We will not do it here because
they are not important to the following discussions.
Notice that the low energy effective gauge potential must
satisfy anti-periodic boundary condition around a vortex.

It is easier to visualize the effect of the instanton in
$3+1$ dimensional spacetime. In $3+1$ dimensions, the
topological defect corresponding to the vortex is the
cosmic string. It can be infinite long or form a close
loop. Just like the vortex, the low energy effective
gauge potential must be anti-periodic around the cosmic
string. If we put our torus in a three dimensional space, we
could say that the anti-periodic boundary condition of the
non-trivial sectors is induced by a cosmic string which goes
through the ``hole'' of the torus. (If the boundary conditions
for both loops on the torus are non-trivial, we
may put a cosmic string through the ``hole,'' and put a
close loop of string ``inside'' the torus. For simplicity,
we consider non-trivial condition on only one boundary.)

To tunnel back to the trivial sector, the cosmic string
must be pulled out of the torus. During the process, the cosmic
string will in general cut the torus at two points. Since
the section of a string is a vortex, the two points are exactly
the positions of a pair of vortex-anti-vortex in the $2+1$
dimensional world. (In the case of $O(2)$, the anti-vortex is
the vortex itself.)

If we restrict ourselves to the torus, what we see about
the pulling out of the cosmic string is the instanton. The
whole process of instanton is the following. If the holonomy
of the $x$-direction is non-trivial, a pair of
vortex-anti-vortex will nucleate, then they wind around the
loop in the $y$-direction and finally, they annihilate each
other. Since the gauge potential is anti-periodic around a
vortex or anti-vortex, after the instanton occurs, the
anti-periodic boundary condition on the $x$-loop will
become periodic.

The mass of the vortex is of order $v^2$ and it must
wind around the non-trivial loop of the torus. Therefore,
we expect that the Euclidean action for the instanton
is in the form $\exp(-{\rm const}\, Lv^2)$.

We can  obtain the same type of  of the instanton action
 considering  the simpliest trial function interpolating
 between two vacua
\begin{equation}
\phi(t)=\frac{t}{T}\phi_1 +\frac{T-t}{T}\phi_2
\end{equation}
where $\phi_1$ and $\phi_2$ are the Higgs fields in these  two
vacua given by (\ref{vacHiggs}), $T$ is the size (in the
time axis) of the instanton and we set the gauge potential to
zero. It is then straight forward to calculate the Euclidean
action for this field configuration. The result is, for example,
\begin{equation}
S_E(\theta=0\to\theta=\frac{\pi x}{L})=
\frac{9L^2v^2}{4T}+(\frac{3}{2}\pi^2 v^2 +w)T
\end{equation}
where $w=\frac{81}{80}L^2v^4(4\lambda_1+v^2\lambda_2)$ is the
contribution from the potential term. Notice that $T$ is of
order $L$ when the Euclidean action is minimum, which is of
order $L v^2$, thus the transition probability will be
suppressed by a factor $\exp(-{\rm const}\, Lv^2)$, as expected.

If the quantum corrections did not lift the degeneracy of the four
classical vacua, there would be mixings between them and the true
ground  state will be some superposition of the classical vacua.
Physical properties of this kind of mixings were considered
earlier  in \cite{kmlee}.   However in this case. as we are
going to argue, the  quantum corrections do lift the degeneracy.

By the analysis of the previous section, in the trivial sector,
the theory is free but the momentum of the effective particle
must be quantized as integral multiple of $2\pi /L$ in both
directions. Thus, the vacuum energy of the trivial sector, which
is one half of the sum of frequencies of all modes, is
\begin{equation}
E_0=\frac{1}{2}\sum_{n,m}((2\pi n/L)^2+(2\pi m/L)^2+M^2)^{1/2}
\end{equation}
Similarly, the vacuum energies of the non-trivial sectors are
\begin{eqnarray}
E_x=E_y&=&\frac{1}{2}\sum_{n,m}
  ((2\pi (n+\frac{1}{2})/L)^2+(2\pi m/L)^2 +M^2)^{1/2}\\
E_{xy}&=&\frac{1}{2}\sum_{n,m}
  ((2\pi (n+\frac{1}{2})/L)^2+(2\pi (m+\frac{1}{2})/L)^2 +M^2)^{1/2}
\end{eqnarray}
We will show that $E_x > E_0$ now. Apart from some unimportant
factor, $E_0=\sum(n^2+m^2+a^2)^{1/2}$ and
$E_x=\sum((n+\frac{1}{2})^2+m^2+a^2)^{1/2}$. Of course,
they diverge. We will use the zeta function regularization.
Consider $E_0(s)=\sum(n^2+m^2+a^2)^{-s}$.
\begin{eqnarray}
E_0(s) \Gamma(s)&=&\sum_{n,m}(n^2+m^2+a^2)^{-s}\Gamma(s)\nonumber\\
  &=&\int^{\infty}_0
     t^{s-1} \sum_{n,m} e^{-(n^2+m^2+a^2)t} dt.
\end{eqnarray}
Similarly,
\begin{equation}
E_x(s) \Gamma(s)=\int^{\infty}_0
     t^{s-1} \sum_{n,m} e^{-((n+\frac{1}{2})^2+m^2+a^2)t} dt.
\end{equation}
The difference is
\begin{eqnarray}
E_x(s)-E_0(s)&=&\frac{-1}{\Gamma(s)}\int^{\infty}_0
     t^{s-1} \sum_{m}e^{-(m^2+a^2)t}
     \sum_n(e^{-n^2t} - e^{-(n+\frac{1}{2})^2 t}) dt\nonumber\\
&=&\frac{-1}{\Gamma(s)}\int^{\infty}_0
     t^{s-1} e^{-a^2 t} \vartheta_1(it/\pi)
     (\vartheta_1(it/\pi)-\vartheta_2(it/\pi)) dt
\end{eqnarray}
where $\vartheta_1(\tau)=\sum\exp{i\pi n^2\tau}$ and
$\vartheta_2(\tau)=\sum\exp{i\pi (n+1/2)^2\tau}$ are some
special cases of the theta functions \cite{mumford}. We need their
properties that
\begin{eqnarray}
\vartheta_1(-1/\tau)&=&(-i\tau)^{1/2}\vartheta_1(\tau)\nonumber\\
\vartheta_2(-1/\tau)&=&(-i\tau)^{1/2}\vartheta_3(\tau)
\end{eqnarray}
where $\vartheta_3(\tau)=\sum(-1)^n\exp{i\pi n^2\tau}$.
For small $t$, $\vartheta_3(i\pi t)$ converges but
$\vartheta_1(i\pi t)$ diverges as $1/\sqrt{\pi t}$. For
large $t$, $\vartheta_1(i\pi t)=1$. We also have
\begin{equation}
\vartheta_1(i\pi t)-\vartheta_3(i\pi t)=\sum_{{\rm odd}\, n}
2 e^{-n^2t} < 4 e^{-t}\sum_n e^{-2nt}=\frac{4 e^{-t}}{1-e^{-2t}}.
\label{difftheta}\end{equation}
Taking all these properties into account, it is easy to see that
\begin{eqnarray}
E_x(s)-E_0(s)&=&\frac{-1}{\Gamma(s)}\int^{\infty}_0
     t^{s-1} e^{-a^2 t} \vartheta_1(it/\pi)
     (\vartheta_1(it/\pi)-\vartheta_2(it/\pi)) dt\nonumber\\
&=&\frac{-\pi}{\Gamma(s)}\int^{\infty}_0
     t^{-s} e^{-a^2/t}\vartheta_1(i\pi t)
     (\vartheta_1(i\pi t)-\vartheta_3(i\pi t)) dt
\end{eqnarray}
converges for $s=-1/2$ for both small and large $t$. By
(\ref{difftheta}), for large $a$, the integrand is in the
form $\exp(1/2 \log t - a^2/t -t)$. The minimum of the
exponent is around $t\approx a=ML/2\pi$. Thus, the splitting
of vacuum energies is of order $\exp(-{\rm const}\,ML)$.
Notice that both the integral and $-\pi/\Gamma(-1/2)=\sqrt\pi/2$
are positive. We have $E_x>E_0$. Similarly, $E_{xy}>E_x$. In
conclusion, there is no mixing between different classical vacua
and the vacuum of the trivial sector is the true vacuum after
quantum correction. The vacua of the non-trivial sectors are
metastable and will decay into the vacuum of the trivial sector
through the vortex type instanton transitions discussed above.
The transition probability is suppressed as $\exp(-{\rm const}\,
L v^2)$, where the action is the double instanton action
corresponding to the bounce solution.

\newsection{Conclusion}
We have discussed the topologically massive gauge theory with
$O(2)$ symmetry on a torus. We found that there are four different
sectors corresponding to the different holonomy on the non-trivial
loops. All four sectors can be described by a free particle of mass
$k\gamma /4\pi$. However, its momentum in the direction of
non-trivial holonomy is restricted to half integral multiple of
some fundamental unit determined by the size of the torus, whereas
the momentum in other direction is whole integral multiple.
In addition to the restriction of momenta, the number of quantum
states of the first Landau level in the trivial sector is only half
of that of the abelian theory. We have also discussed the transitions
between different vacua induced by the instantons. A very clear
picture on effect of the instanton in term of cosmic string in
$3+1$ dimensions is given. The transition probability is
estimated to be suppressed by an exponential factor
proportional to the size of the torus. After including the
quantum effect, the true vacuum is the vacuum of the trivial
sector and other classical vacua are metastable.

\medskip

\noindent
{\bf Acknowledgments.}

\noindent
We would like to thank Leith Cooper for pointing out the typos in
Section 2. This work was supported in part by the PPARC
grant GR/J 21354.

\renewcommand{\Large}{\normalsize}

\end{document}